\documentclass[a4paper,11pt]{article}
\usepackage[a4paper, total={17cm, 25cm}]{geometry}
\usepackage{fancyhdr}
\usepackage{graphicx}
\usepackage{subcaption}
\fancypagestyle{plain}{
\fancyhf{}
\fancyhead[RH]{\thepage}
\fancyfoot[CF]{28th Texas Symposium on Relativistic Astrophysics \\ Geneva, Switzerland -- December 13-18, 2015}

}
\begin{document}
%%%%%%%%%%%%%%%%%%%%%%%%%%%%%%%%%%%%%%%%%%%%%%%%%%%%%%%%%%%%%%%%%%%%%%%%%%%%
%%%% Title
%%Latex Turkish Characters: \u{g} – ğ; \u{G} – Ğ; \c{c} – ç; \c{C} – Ç; \c{s} – ş; \c{S} – Ş; \”{u} – ü; \”{U} – Ü
\title{\bf Binary Systems with a Black Hole Component as Sources of Gravitational Waves}

\author{
Dolunay Ko\c{c}ak$^1$ and Kadri Yakut$^1$ \\
$^1$ University of Ege, Faculty of Science, Department of Astronomy and Space Sciences, \.Izmir, Turkey}

\date{February 15, 2016}

\maketitle

\begin{abstract}
Discovery of gravitational waves by LIGO team \cite{LIGO2016a} bring a new era for observation of black hole systems.
These new observations will improve our knowledge on black holes and gravitational physics. In this study,
we present angular momentum loss mechanism through gravitational radiation for selected X-ray binary systems.
The angular momentum loss in X-ray binary systems with a black hole companion due to gravitational radiation and mass loss
time-scales are estimated for each selected system. In addition, their gravitational wave amplitudes are also estimated and
their detectability with gravitational wave detectors has been discussed.

\end{abstract}
%%%%%%%%%%%%%%%%%%%%%%%%%%%%%%%%%%%%%%%%%%%%%%%%%%%%%%%%%%%%%%%%%%%%%%%%%%%%

%%%% Paper body

\section{Introduction}

In astrophysics, binary stars provide an excelent laboratory to study stellar properties.
Analysis of a double-lined spectroscopic binary with an eclipsing light variation gives information about mass, radius,
luminosities and the distance of the system. Most of the stars whose masses and other physical parameters are known,
are member of binary systems. It has been discovered observationally Cyg X-1 to have a black hole companion,
because of its binary property \cite{Bowyer65}.

Massive stars at the end of their evolution following a core collapse/supernova, they leave behind either a neutron
star or a black hole. The processes may differ in a single star and in a star that is a member of a binary system.
In binary systems the massive component evolve faster to form an X-ray binary system. One of the component in an X-ray
binary system is a black hole or a neutron star. X-ray binaries can be divided into three subclasses according to the masses of
the companion stars: (i) low-mass X-ray binary (LMXB), (ii) intermediate-mass X-ray binary (IMXB),  (iii) high-mass X-ray binary (HMXB).
After discovery of gravitational wave event by LIGO team (Abbott et al. 2006) significance of compact binary systems is increased.
Especially those close binary systems with a black hole components, since they could be the progenitor of binary black hole systems and therefore candidates for gravitational wave astronomy.
In this project, X-ray binary systems with black hole components have been studied. Physical and orbital parameters of the
binary systems have been collected from the literature to study their long-term light variations and angular momentum loss mechanisms.

\section{Angular Momentum Loss Mechanisms in Binary Star Systems}
Angular momentum loss mechanisms in a binary system depend on the system’s type and its geometric structure.
For instance, in main sequence stars angular momentum loss via stellar wind is dominant (see \cite{Yakut05}, \cite{Yakut08}, \cite{Kawaler98})
while in cataclysmic variables and in some other X-ray binaries both mass loss by stellar activity and angular momentum via gravitational
radiation is important (see \cite{Kalomeni10}, \cite{Andronov03} and references therein).
If both stars are compact and the orbital period is short enough, gravitational radiation is the main mechanism of angular momentum loss.
One of the predictions of the general relativity is the energy loss caused by perturbation in space-time
of binary stars that releases gravitational waves \cite{EInstein16a}, \cite{EInstein16b}, \cite{EInstein18}.

\section{Angular Momentum Loss Mechanisms in Binary Systems with a Black Hole Components}

In this study, observational results of 27 binary systems with black hole components have been catalogued.
Orbital period of these systems vary from 0.2 to 34 days while black hole masses vary from 4 to 23 solar masses.
Most of the binary systems have circular orbits (e=0) while a few of them have eccentric orbits.
In X-ray binary systems, mass loss via both late type and early type stellar activities are observed.
This is even more important in systems where the separation of the components decreases.
Angular momentum loss via gravitational radiation present in addition to the angular momentum loss via mass loss.
Angular momentum loss time scale via stellar wind and gravitational radiation are given by \cite{Yakut08} as

\begin{equation}
\tau_{MSW} = 14 (\frac{M}{M_\odot})^{2/3} (\frac{R_2}{R_\odot})^{-4}(\frac{P}{\rm{day}})^{10/3}(1+q)^{-1}(1-e^2)^{1/2} \rm{Gyr}.\label{Eq-timescale-msw}
\end{equation}

\begin{equation}
\tau_{GR} = 376.4q^{-1}(1+q)^2M^{-5/3}P^{8/3} (1-e^2)^{3/2}(1+\frac{7}{4}e^2)^{-1}  \textrm{Gyr}. \label{Eq-timescale-gr}
\end{equation}

Angular momentum loss rates and time scales of the selected systems are estimated by using Eq.~(1) and Eq.~(2).
The results are shown in Fig. 1 and Fig. 2.

\begin{figure}
\centering
\begin{minipage}{0.49\textwidth}
\centering
\includegraphics[width=.9\textwidth]{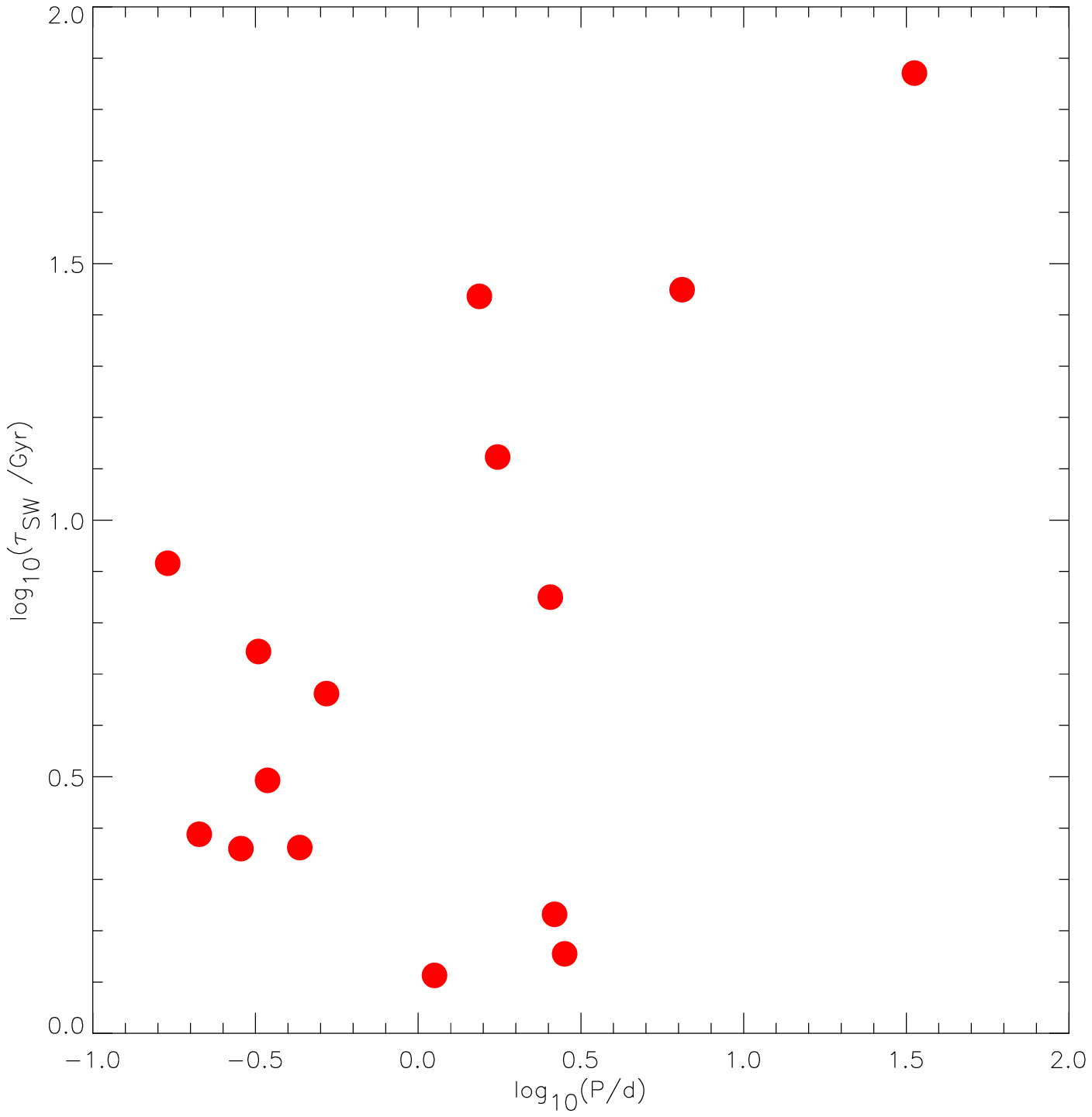}
\caption{Plot of  $\log \tau_{\textrm{MSW}}$ {\it vs.} $\log$P for binary systems with black hole components.}\label{fig1}
\end{minipage}
\begin{minipage}{0.49\textwidth}
\centering
\includegraphics[width=.9\textwidth]{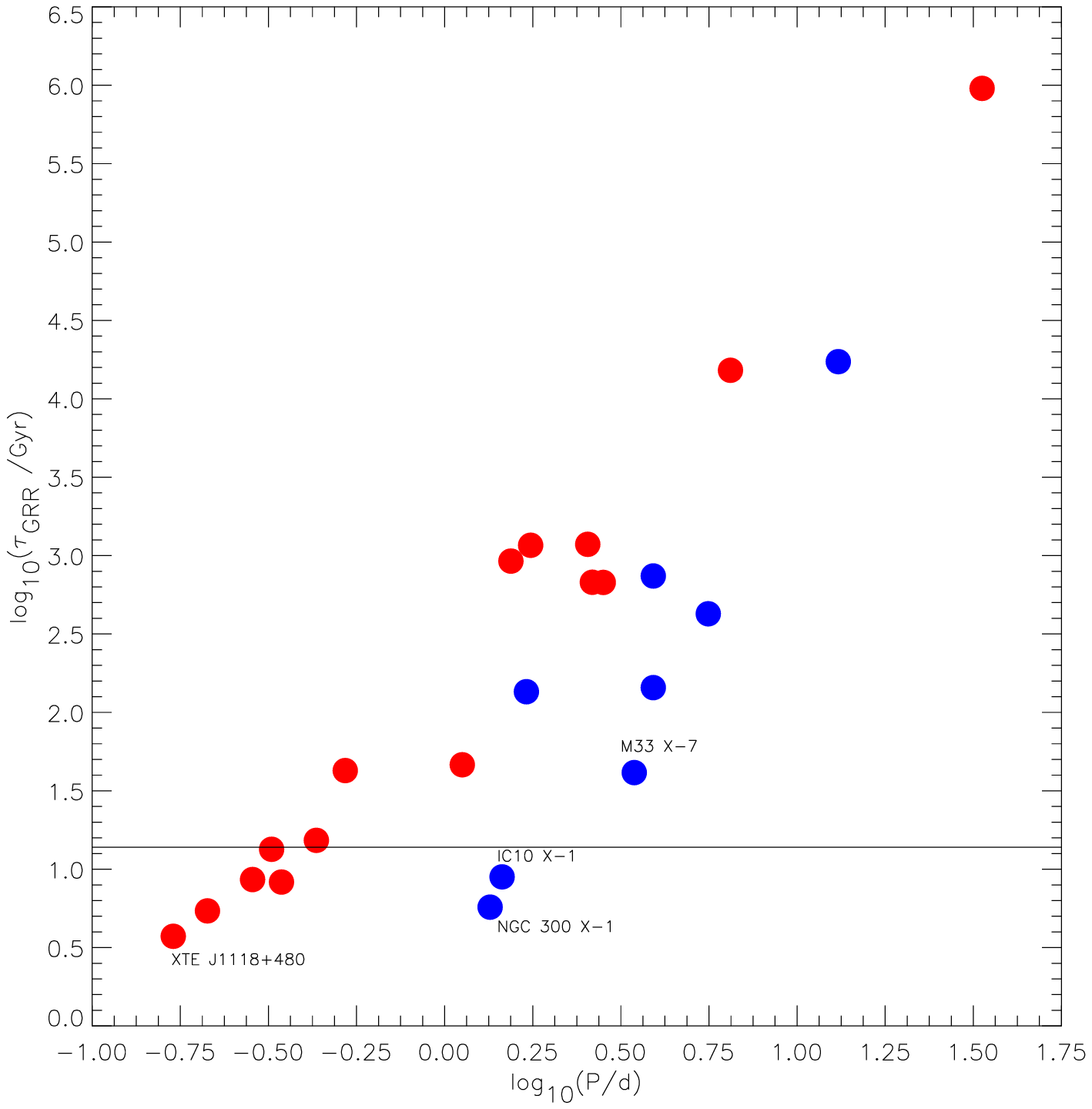}
\caption{Plot of $\log \tau_{\textrm{GR}}$ {\it vs.} $\log$P for binary systems with black hole components.}\label{fig2}
\label{fig:right}
\end{minipage}
\end{figure}

\begin{figure}
\includegraphics[width=.6\textwidth]{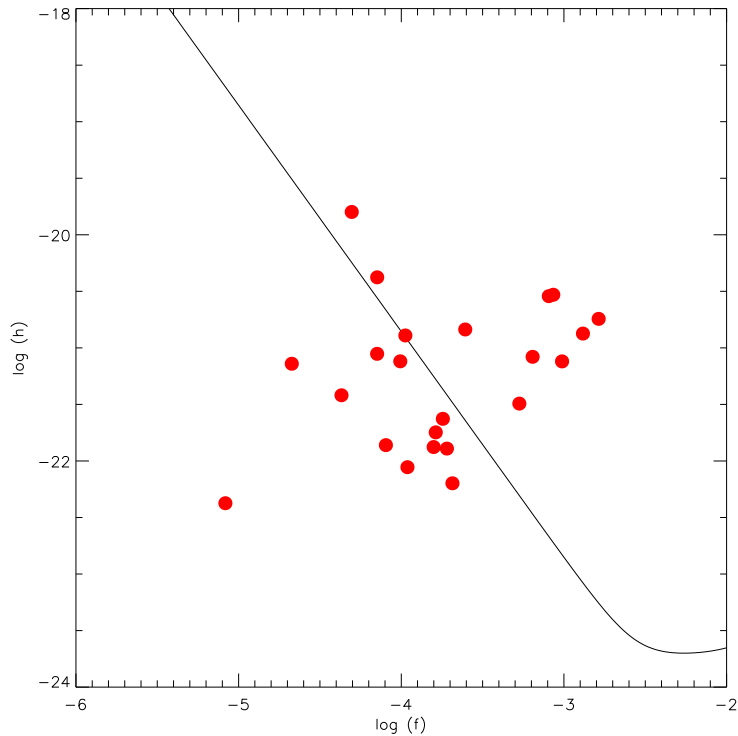}
\caption{Plot of $\log h$ {\it vs.} $\log f$ for binary systems with black hole components. Continuous line shows the expected sensitivity of LISA.}\label{fig3}
\end{figure}

Binary systems consisting of black hole components with short orbital periods are candidates for gravitational-wave sources.
GW observation is a tools for detecting double black hole binaries, which is already observed by LIGO \cite{LIGO2016a}.
GW amplitude (h) of these systems are compared with the limit of gravitational wave interferometers (e.g. LISA, LIGO, VIRGO).
Starting from the linearized Einstein field equation one can calculate the GW amplitude (h) as is given in Eq. (3) \cite{OY2012}. In Eq. (3) f is the frequency,
D is the distance of source in Mpc, and M is the total mass in solar unit.
The gravitational wave amplitudes of the selected systems are shown in Fig. 3.
In the figure we plotted selected binary systems with black holes within the observation limit of LISA satellite.

\begin{equation}
h \simeq 2.5 \times 10^{-22} {M_{1}M_{2}}{M^{-1/3}D^{-1}}f^{2/3} \label{amplitude}
\end{equation}

\textbf{Acknowledgments}\\
This study was supported by the Turkish Scientific and Research Council (T\"UB\.ITAK 113F097).
The current study is a part of MSc thesis by D. Ko\c{c}ak.
\paragraph{}
%%%% References

%%%%%%%%%%%%%%%%%%%%%%%%%%%%%%%%%%%%%%%%%%%%%%%%%%%%%%%%%%%%%%%%%%%%%%%%%%%%
\end{document}